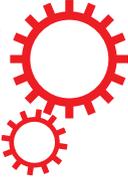

# SCIENTIFIC REPORTS

**OPEN**

# A Geometric-Structure Theory for Maximally Random Jammed Packings

Jianxiang Tian[1,2], Yaopengxiao Xu[3], Yang Jiao[3] & Salvatore Torquato[4,5,6]

**Maximally random jammed (MRJ) particle packings can be viewed as prototypical glasses in that they are maximally disordered while simultaneously being mechanically rigid. The prediction of the MRJ packing density $\phi_{MRJ}$, among other packing properties of frictionless particles, still poses many theoretical challenges, even for congruent spheres or disks. Using the geometric-structure approach, we derive for the first time a highly accurate formula for MRJ densities for a very wide class of two-dimensional frictionless packings, namely, binary convex superdisks, with shapes that continuously interpolate between circles and squares. By incorporating specific attributes of MRJ states and a novel organizing principle, our formula yields predictions of $\phi_{MRJ}$ that are in excellent agreement with corresponding computer-simulation estimates in almost the entire $\alpha$-$x$ plane with semi-axis ratio $\alpha$ and small-particle relative number concentration $x$. Importantly, in the monodisperse circle limit, the predicted $\phi_{MRJ} = 0.834$ agrees very well with the very recently numerically discovered MRJ density of 0.827, which distinguishes it from high-density "random-close packing" polycrystalline states and hence provides a stringent test on the theory. Similarly, for non-circular monodisperse superdisks, we predict MRJ states with densities that are appreciably smaller than is conventionally thought to be achievable by standard packing protocols.**

A packing is defined as a large collection of nonoverlapping solid objects in $d$-dimensional Euclidean space $\mathbb{R}^d$. A simple but important attribute of a packing is its packing density $\phi$, defined as the fraction of space $\mathbb{R}^d$ covered by the particles. Dense particle packings are useful models of a variety of equilibrium and non-equilibrium low-temperature states of matter[1–4], granular media[2,5–9], heterogeneous materials[2], and biological systems[10,11].

In general, for a given particle shape, the associated packings can possess a diversity of densities and degrees of order[12,13]. Maximally dense packings of frictionless congruent particles, usually achieved by ordered particle configurations, are the thermodynamically stable phases in the infinite-pressure limit and hence also determine the high-density equilibrium phase behavior of such hard-particle systems[13].

On the other hand, the maximally random jammed (MRJ) packings of frictionless hard particles are nonequilibrium states that can be viewed as prototypical glasses in that they are maximally disordered while simultaneously being perfectly mechanically rigid[12,13]. It is important to note that the concept of MRJ is geometric-structure based, which emphasizes the analysis of individual packing configurations regardless of their probability of occurrence[12,13]. Therefore, MRJ states are distinct from the so-called random close packed (RCP) states[14], which have recently been suggested to be the most probable jammed configurations within an ensemble[15]. While RCP and MRJ packings are believed to have similar densities in $\mathbb{R}^3$, other attributes can be distinctly different[13,16], and their packing characteristics can be dramatically different from one another in two dimensions, including the respective densities[17]. MRJ packings possess the singular property of hyperuniformity[18,19], i.e., infinite-wavelength density (local-volume-fraction) fluctuations are anomalously suppressed. In MRJ packings, this is manifested as negative quasi-long-range (QLR) pair correlations that decay asymptotically like $-1/r^{(d+1)}$[20–22]. Such large-scale "hidden" order that is not manifested on short length scales has also been identified in a number of other physical systems[23–27] and, more recently, in avian photoreceptor patterns[28]. It is also noteworthy that a variety of other hyperuniform particle systems with different long-range correlations have been identified[29–34].

[1]Department of Physics, Qufu Normal University, Qufu 273165, China. [2]Department of Physics, Dalian University of Technology, Dalian 116024, China. [3]Materials Science and Engineering, Arizona State University, Tempe Arizona 85287, USA. [4]Department of Chemistry, Princeton University, Princeton New Jersey 08544, USA. [5]Department of Physics, Princeton University, Princeton New Jersey 08544, USA. [6]Program in Applied and Computational Mathematics, Princeton University, Princeton New Jersey 08544, USA. Correspondence and requests for materials should be addressed to J.T. (email: jxtian@dlut.edu.cn) or S.T. (email: torquato@princeton.edu)

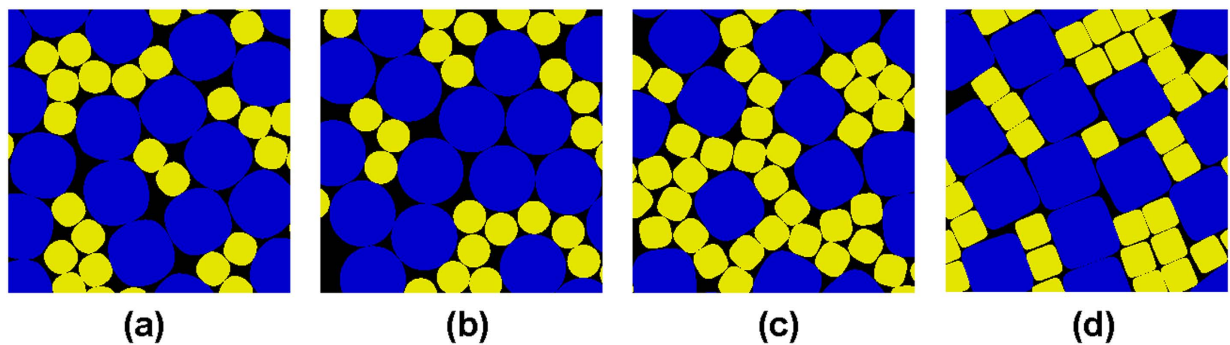

Figure 1. **Portions of distinct types of MRJ packings of binary superdisks with semi-axis ratio $\alpha = 0.5$ and small-particle relative number concentration $x = 0.5$.** (**a**) $p = 0.85$. (**b**) $p = 1.0$. (**c**) $p = 1.5$. (**d**) $p = 3.0$.

Predicting the packing density $\phi$ of jammed disordered packings as a function of particle characteristics, such as particle shape and size distributions, is a long-standing and notoriously difficult problem. This is because the final states depend on many factors, such as the preparation technique, degree of order and interparticle friction, to mention just a few[12,13,35]. The determination of the maximal packing density $\phi_{max}$ for non-space-filling particles has been a source of fascination for centuries[36,37]. Recently, general organizing principles and conjectures have been proposed that enable one to predict dense packing arrangements of frictionless nonspherical particles based on the geometrical properties of the shapes[38–41]. This allowed the prediction of $\phi_{max}$ of multiply connected nonconvex bodies such as tori[42]. The estimation of the MRJ density $\phi_{MRJ}$ is even more challenging. Whereas predicting $\phi_{max}$ only requires the consideration of relatively simple ordered arrangements of a small number of particles in a periodic fundamental cell, to predict $\phi_{MRJ}$ it is necessary to capture complex structural features of very large disordered packings that simultaneously possess strong short-range and hyperuniform long-range correlations[13,18,20–22].

A number of schemes to predict RCP densities have been developed. Recently, a 3D "granocentric" model for random packings of colloidal spheres has been devised, which distills the complexity of global packing into local stochastic processes and yields accurate estimates of the overall packing density for a variety of particle size distributions[43]. Moreover, a mean-field theory for RCP based on Edwards' ensemble has been developed, which leads to a predictive framework to calculate packing density for a wide class of particles including both spherical and non-spherical shapes (e.g., ellipsoids, spherocylinders and polyhedra) based on the distribution of Voronoi volumes associated with a particle[44,45].

An important family of two-dimensional nonspherical shapes are given by superdisks. A superdisk is a centrally symmetric body defined by

$$|x_1|^{2p} + |x_2|^{2p} \leq 1, \qquad (1)$$

where $x_1$ and $x_2$ are Cartesian coordinates and $p \in [0, \infty]$ is the deformation parameter, which indicates to what extent the particle shape has deformed from that of a circular disk ($p = 1$). As $p$ continuously increases from unity to infinity, a superdisk continuously changes its shape from a circle to a square. As $p$ decreases from 1 to 1/2, another family of square-symmetric superdisks is obtained, but with the symmetry axes rotated 45 degrees with respect to that of the first family. When $p < 1/2$, the superdisk is concave and becomes a cross in the limit $p \rightarrow 0$.

In this paper, using the geometric-structure approach[13], we derive for the first time a highly accurate formula for the densities of MRJ packings of frictionless two-dimensional (2D) binary (two-component) convex superdisks, which constitutes a family of an uncountably infinite number of distinct types of packings (both in particle shape and relative concentrations, see Fig. 1). Our formula, which is based on a novel organizing principle for such MRJ packings derived here, explicitly incorporates *specific attributes* of MRJ states, including jamming induced local spatial correlations and hyperuniformity, and thus, is distinctly different from the aforementioned schemes[43–45]. Our formula for frictionless binary superdisks is in excellent agreement with computer simulations (see Methods) for the entire $\alpha$-$x$ plane with semi-axis ratio $\alpha$ and small-particle relative number concentration $x$, and is easily generalized to other smooth non-spherical shapes and other dimensions.

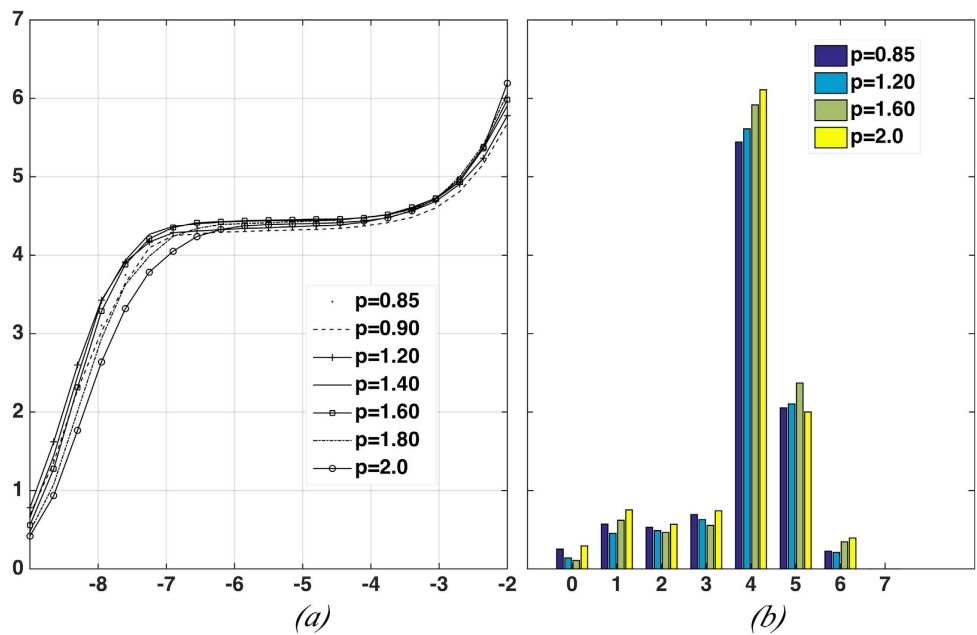

**Figure 2.** (**a**) Average number of contacts per particle as a function of interparticle gap tolerance for various MRJ superdisk packings. (**b**) Distribution of the number of contacts per particle for different packings.

It is noteworthy that for the special limit of monodisperse circular disks ($p = \alpha = x = 1$), our prediction $\phi_{MRJ} = 0.834$ is in very good agreement with the recently numerically discovered MRJ isostatic state with a density of $\phi_{MRJ} = 0.827$[17]. The latter is a significant development because it was previously thought that either RCP states for monodisperse disks did not exist (because standard numerical protocols typically produce highly ordered, polycrystalline packings due to a lack of geometrical frustration) or, by entropic measures, RCP states were indeed these highly ordered polycrystalline arrangements. The fact that our prediction for $\phi_{MRJ}$ in this instance is consistent with this new data is a testament to the power of the approach. Moreover, for most monodisperse superdisks that are not circles, our predicted MRJ densities are significantly lower than packing densities of polycrystalline packings that would be achieved by standard numerical and experimental packing protocols.

## Results

**Novel organizing principle for MRJ packings.** The numerically generated high-quality strictly jammed packing configurations allow us to obtain robust statistics of the interparticle contacts. Specifically, we compute the average number of contacts per particle as a function of interparticle gap tolerance, which possesses a plateau over a wide range of tolerance values (see Fig. 2a). This feature enables us to determine the contact neighbors of a particle and to locate the contact points to a heretofore unachieved accuracy. The distributions of the number of contacts per particle for different packings are shown in Fig. 2b. Note that in these packings, the total number of contacts (i.e., constraints) $N_c$, which can be easily computed from the contact number distributions, is smaller than the total number of degrees of freedom (DOF) $N_f = 3N + 3$, which includes the contribution from the 2 translational DOF and 1 rotational DOF- associated with the $N$ particles in the packing as well as the 3 DOF of the deformable boundary[16]. Unlike MRJ packings of circular disks, which are isostatic (i.e., $N_c = N_f$), MRJ superdisk packings are *hypostatic* (i.e., with $N_c < N_f$ yet strictly jammed). This indicates that the orientations of the contacting superdisks are necessarily correlated to block the relative rotations of the particles, which play a trivial role in MRJ packings of circular disks[46].

We now devise an organizing principle by analyzing hypostatic contacting particle configurations. We observe that in MRJ packings, the contact points on a particle associated with small surface curvatures are favored over those associated with large surface curvatures. This feature is explained by the fact that the small-curvature contacts more efficiently block particle rotations[46,47]. However, the value of surface curvature depends on both particle shape and size, and thus, is different for superdisks with different $p$ and $\alpha$ values. Therefore, we examine contact angles instead of contact curvatures for our analysis. Specifically, a contact angle is defined as the angle between the line connecting the particle center to a contact point on particle surface and the horizontal principal axis when the particle is aligned with the axes of a local Cartesian coordinate system. For a specific $p$ value, the local particle surface curvature $\kappa$ at the contact point is related to the contact angle $\theta$ via:

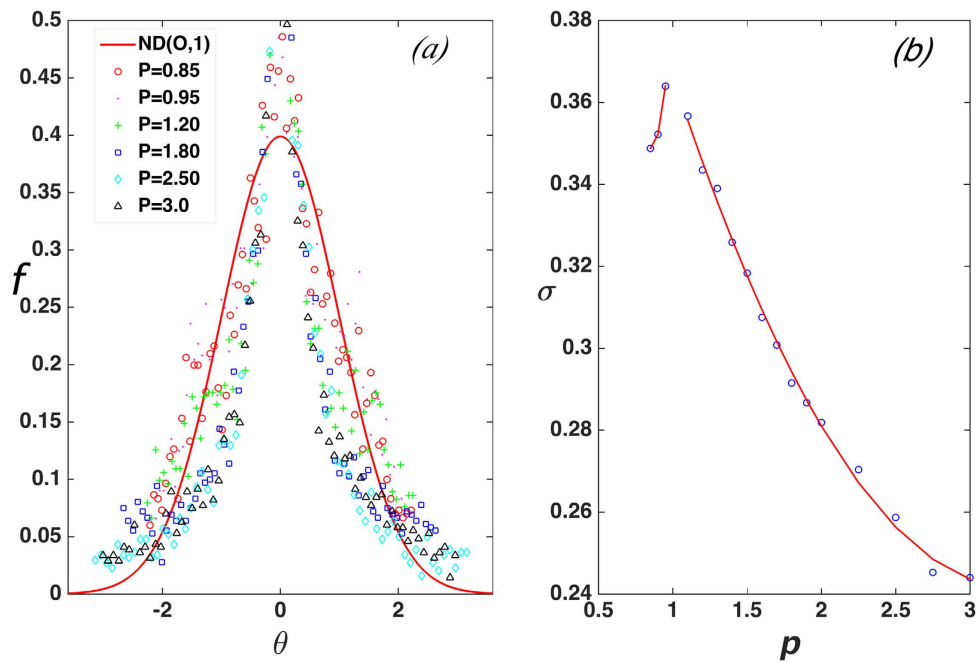

**Figure 3.** (**a**) Rescaled distributions of contact angle for superdisks with different deformation parameter $p$ (i.e., particle shape). Also shown is a universal Gaussian curve on which all data collapse onto. (**b**) Standard deviations of the contact angle distributions associated with different particle shape (i.e., different $p$ values).

$$\kappa = \frac{(2p-1)(\cos\theta \ \sin\theta)^{\frac{2p+1}{p}}}{R\left\{(\cos\theta)^{\frac{2}{p}}(\sin\theta)^4 + (\sin\theta)^{\frac{2}{p}}(\cos\theta)^4\right\}^{\frac{3}{2}}}, \quad (2)$$

where $\theta \in (0, \pi/2]$. Because the particle possess 4-fold rotational symmetry, the curvature value associated with other angles can be obtained by translating the angles to the interval $(0, \pi/2]$. Since small-curvature contacts are more efficient in blocking relative particle rotations, it is reasonable to assume that for any specific $p$ value, the most probable contact angle $\theta^*$ (i.e., the mean of the distribution) corresponds to the minimal curvature $\kappa_{min}$ associated with the particle shape. Due to the symmetry of the particle shape, we expect the distribution of contact angles $f(\theta)$ is also symmetric about $\theta^*$. We thus propose that $f(\theta)$ possesses a Gaussian form, i.e.,

$$f(\theta) = \frac{1}{\sigma\sqrt{2\pi}} \exp\left[-\frac{(\theta-\theta^*)}{2\sigma^2}\right]. \quad (3)$$

where the mean $\theta^*$ and standard deviation $\sigma$ depend alone on the particle shape (i.e., the deformation parameter $p$).

To validate (3), we compute the statistics of contact angles from the numerically generated packing configurations. For each particle in a MRJ packing, the locations of contact points on the particle surface are identified and the associated contact angles are computed. The contact angles for all particles in the packing are then binned to generate a histogram, or a distribution of the angles. We test the validity of the distribution using two different methods (i.e., Kolmogorov-Smirnov test and t-test) and find that for any given particle shape that we considered, the distribution of contact angles indeed possesses the Gaussian form (3). In addition, after simple re-scaling, the contact angle distributions for different particle shapes all collapse onto a universal standard Gaussian distribution curve with zero mean and unitary standard deviation, as shown in Fig. 3a. We note that the original mean value of contact angle for superdisks with $p < 1$ is $\pi/4$ and the mean for superdisks with $p > 1$ is $\pi/2$. Figure 3b shows the standard deviations of the un-scaled contact angle distributions as a function of $p$. This collapse strongly suggests the existence of a universal organizing principle for MRJ packings of frictionless superdisks independent of the detailed particle shape, which we elaborate on below.

As shown in Fig. 3b, the standard deviation $\sigma$ is a monotonic decreasing function of $|p-1|$ and possesses a singularity at $p=1$ (i.e., the circle limit). This indicates that as the shape deviates from that of a circle (with uniform local curvature), the contacts are more likely to occur in low curvature regions on the shape, which results in significant orientational correlation in the packing. For nonspherical shapes, the requirement of jamming in the packing (i.e., mechanical stability) necessarily leads to correlations among the contacts, which block both the translational and rotational motions of any particles in the packing. The observed correlations among contacts, which are quantitatively characterized by the universal Gaussian distribution of contact angles, are sufficient and necessary to induce jamming and thus, are an attribute of MRJ states. Therefore, the capability of blocking relative particle motions via contacts at different local surface curvatures associated with different contact angles does not depend on the specific particle shape and spatial dimension. This is not only a new organizing principle for MRJ packings of frictionless superdisks but other frictionless nonspherical particles with a sufficiently smooth shape (e.g., ellipses). A quantitative consequence of this organizing principle is the universal Gaussian distribution of contact angles. Moreover, an accurate analytical formalism for predicting $\phi_{MRJ}$ should quantitatively incorporate such jamming induced spatial correlations, as we do here.

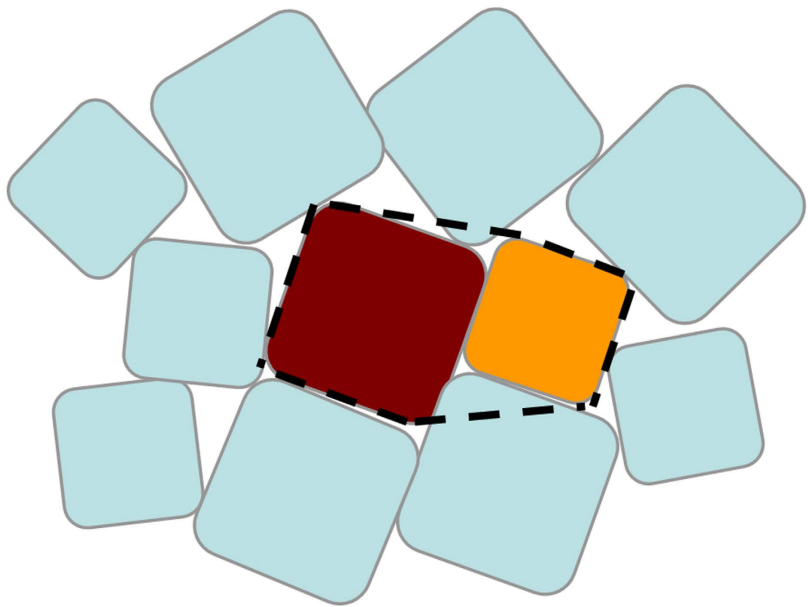

**Figure 4.** A schematic illustration of a local two-particle packing configuration employed to estimate the global MRJ packing density.

**Accurate prediction of MRJ packing densities.** We now employ our organizing principle to estimate the MRJ packing density $\phi_{MRJ}$ for frictionless superdisk systems for various values of $p$, $x$ and $\alpha$. The salient idea is to approximate the global packing density by a contact-configuration averaged local packing density. We first provide arguments why such a local analysis can lead to accurate predictions of $\phi_{MRJ}$. We note that the MRJ packings are hyperuniform with long-range correlations[13,18]. A unique feature of these hyperuniform systems is that the local-volume-fraction fluctuations decay faster than the inverse of the volume of a spherical observation window of radius $R$, i.e., faster than $R^{-d}$ [21]. This indicates that the local packing density $\tau(R)$ converges to the global packing density $\phi_{MRJ}$ rapidly as the size of the window increases, i.e, effectively $\sigma_\tau^2 = \langle \tau^2(R) \rangle - \phi_{MRJ}^2 \sim R^{-(d+1)}$. In other words, the long-range hyperuniformity property imposes strong suppression of density fluctuations at short scales. This rapid convergence property of disordered hyperuniform jammed packings is supported by previous theoretical studies[13,18,21] as well as experimental investigations[27]. Moreover, to further reinforce this point, we have computed here the variance $\sigma_\tau^2(R)$ of a MRJ packing of binary superdisks with $R$ equal to the diameter of the large particles, and compared it to that of a corresponding saturated non-hyperuniform packing. This non-hyperuniform packing is generated by first randomly and sequentially placing particles with random orientations in a simulation domain without overlapping existing particles, i.e., via the random sequential addition (RSA) process, which leads to a nearly hyperuniform packing[48]. Then the RSA configuration of superdisks is compressed to the densest possible state while fixing the particle orientations. The variance of the MRJ packing in this small window is four times smaller than that of the compressed RSA packing. Therefore, although only first-neighbor shell is considered in our analysis, the organizing principle is sufficient to yield accurate prediction of $\phi_{MRJ}$.

Our density-prediction procedure works as follows: we first construct a local two-particle packing configuration, which is a region $\Omega$ enclosed by two line segments that are tangential to two contacting particles and portions of particle surfaces, as shown by the thick dash lines in Fig. 4. We note that this construction allows us to quantitatively incorporate the jamming-induced contact distribution, which encodes explicit two-body statistics. Our approach is to be contrasted with the "granocentric" approach[43] as well as the Edwards' ensemble-approach based on the Voronoi volume distribution[44,45], which is a one-body statistic[49]. In the "granocentric" approach, which was originally devised for 3D colloidal spheres, the formation of an interparticle contact was considered independent of other contacts. However, extending this assumption to 2D is problematic because of the lack of intrinsic geometric frustrations (intrinsic to 3D packings), a consequence of which are substantial correlations between contacting particles. For the Edwards' ensemble-approach, although the Voronoi volume of a particle depends on the centroid positions of its contacting neighbors, unlike in our formalism, the spatial correlations between the central particles and its neighbors were not explicitly incorporated. We then compute the local packing density $\phi_{local} = v_{particle}/v_{local}$, where $v_{particle}$ is the total volume of the two particles and $v_{local}$ is the volume of the region $\Omega$. In other words, we can consider that the 2D plane is approximately tessellated into polygons, each effectively containing two particles.

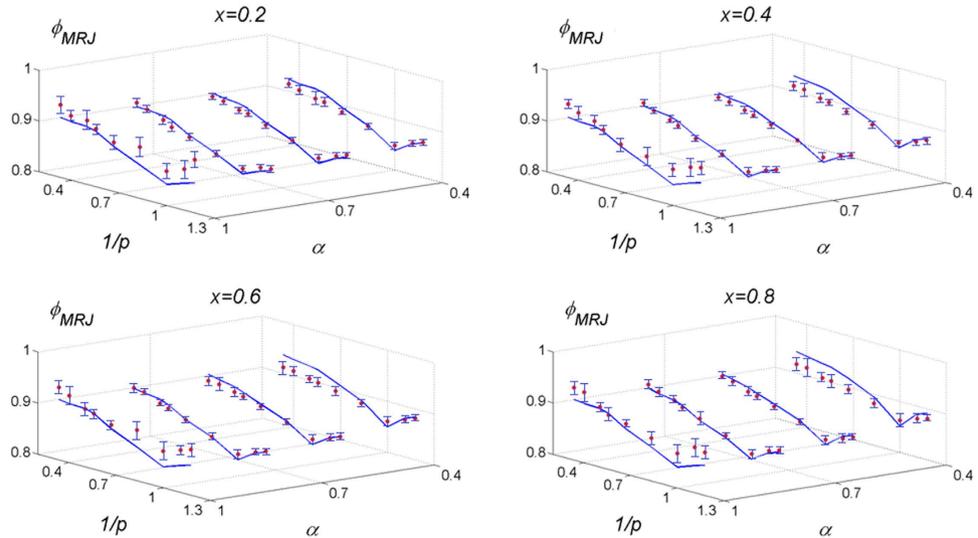

**Figure 5. Comparison of the MRJ packing density estimated using Eq. (4) and obtained from simulations for various binary superdisk packings.** For each panel, the packing density is shown as a function of small-to-large semi axis ratio $\alpha$ and deformation parameter $p$ with a fixed small-particle relative concentration $x$. The predicted values of $\phi_{MRJ}$ are shown as blue curves and the simulated values are shown as red dots with confidence intervals shown.

In general, $\phi_{local}$ depends on the location of contact on the particles, which is characterized by the contact angle $\theta$, the relative orientation of the particles $\Theta$ and size ratio $\alpha$ between the two contacting particles. For a binary packing (the focus of this letter), there are two classes of local contact configurations, i.e, the two particle are of the same size (whose local packing density is denoted by $\phi^{(1)}_{local}$) and two particles with different sizes (with local packing density denoted by $\phi^{(2)}_{local}$). Therefore, the global MRJ density $\phi_{MRJ}$ can be estimated as follows:

$$\phi_{MRJ} = \int_{\theta_1}^{\theta_2} [(1-x)\overline{\phi}^{(1)}_{local}(\theta) + x\overline{\phi}^{(2)}_{local}(\theta)]f(\theta)d\theta \qquad (4)$$

where $\overline{\phi}^{(i)}_{local}$ ($i=1,2$) is the average local density over $\Theta$ at fixed $\alpha$. For $p<1$, $\theta_1=0$ and $\theta_2=\pi/2$; for $p>1$, $\theta_1=\pi/4$ and $\theta_2=3\pi/4$. More generally, formula (14) is easily extended to the case of a polydisperse packing with particle size distribution $P(R)$ (where $R$ is the semi-axis of the superdisks) as follows:

$$\phi_{MRJ} = \int_R \int_{\theta_1}^{\theta_2} \overline{\phi}_{local}(\theta;R)f(\theta)P(R)d\theta dR \qquad (5)$$

where $\overline{\phi}_{local}(\theta;R)$ is the average local density over $\Theta$ and $\alpha$ for a given superdisk with semi-axis $R$.

To verify its accuracy, Eq. (4) is employed to estimate the $\phi_{MRJ}$ for a variety of binary frictionless superdisk packings with $x \in [0.05, 0.95]$, $\alpha \in [0.2, 0.95]$ and $p \in [0.85, 4.5]$. We find that for all values of $\alpha$ and $x$ considered the estimated $\phi_{MRJ}$ always agrees very well with the corresponding simulation results, with the largest deviations smaller than 1.5%, as shown in Fig. 5 for selected values of $x$, $\alpha$ and $p$ (see the Supplementary Information for the values of $\phi_{MRJ}$). For the extreme case where $p=0.5$ and $\infty$, the particle becomes a square with singular curvatures at the corners and our formalism does not hold. In addition, when $\alpha$ is very small (e.g., less than 0.2), even with $x \in [0.05, 0.95]$, the large particles can form a jammed backbone with small particles moving freely within cases formed by large particles[50], and thus, the organizing principle does not hold for these cases.

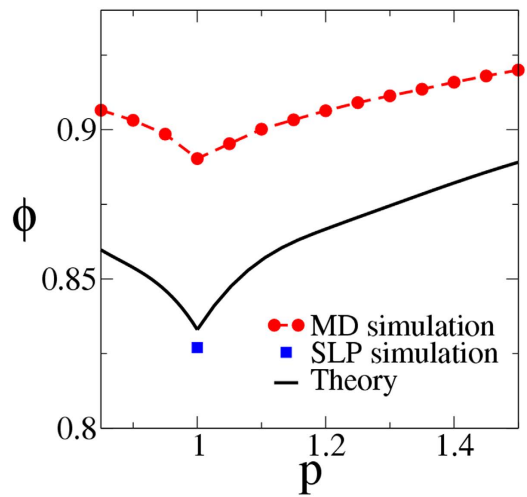

**Figure 6.** Comparison of the MRJ packing density for monodisperse superdisks estimated using Eq. (4) (solid line) and obtained from DTS molecular dynamics (MD) simulations (red circles), which fail to represent MRJ states, as well as sequential linear programming (SLP) datum for the circle case (blue square)[17].

### Discussion
When the particles are nearly identical in size ($\alpha \sim 1$) or when the concentration of one species is overwhelmingly larger than the other ($x \sim 0$ or 1), the most "disordered" packings generated using typical packing simulation protocols, including the one employed here (i.e., DTS molecular dynamics method), possess polycrystalline regions and thus, a high degree of order. This calls into question whether our simulated packings, in these special cases, represent the actual MRJ state as well as the effectiveness of numerical packing protocols in generating true MRJ states. In ref. 16, high-fidelity MRJ *isostatic* packings of equal-sized circular disks were constructed for the first time using a novel sequential linear programming (SLP) packing protocol[51]. Such packings were shown to possess no crystalline regions and an average packing density $\phi_{MRJ} = 0.827$. This singular result provides a stringent benchmark for our predictions of $\phi_{MRJ}$ for equal-sized superdisks. Specifically, in the limit $\alpha = 1$ and $p = 1$, our prediction yields $\phi_{MRJ} = 0.834$, which agrees very well with the numerical result reported in ref. 16. This consistency strongly indicates the accuracy of our formula for the circle case and it is reasonable to expect it yields accurate predictions of $\phi_{MRJ}$ for equal-sized superdisks with $p \neq 1$. Indeed, as shown in Fig. 6, the predicted $\phi_{MRJ}$ for these shapes are at least 5% lower than the corresponding polycrystalline packings produced using the DTS simulations. This provides strong evidence that such polycrystalline packings of superdisks with $p \neq 1$ do not represent the actual MRJ states, and our predicted $\phi_{MRJ}$ correspond to MRJ packing arrangements that are yet to be constructed via certain novel numerical and experimental protocols.

We expect that our formalism yields accurate predictions for $\phi_{MRJ}$ for frictionless superdisks with $x \in [0, 1]$, $\alpha \in [0.2, 1]$ and $p \in (0.65, 5)$, which are beyond the parameter values studied in our numerical simulations. The reason is that it explicitly incorporates characteristics of MRJ states, including hyperuniformity- and jamming-induced local correlations between contacting neighbors, which are crucial ingredients in a novel local organizing principle we report here. Moreover, the derivation based on the novel organizing principle does not depend on specific particle shape (i.e., $p$ values) and size distributions. It is notable that our analysis provides a clear systematic procedure to predict MRJ densities of other families of smooth particle shapes in both 2D and 3D, e.g., ellipses and ellipsoids. We note that after initial submission of our paper, we learned of a preprint that uses a local perturbation analysis to predict random close packing densities of the special case of smooth particle shapes that are very nearly spherical, which can continuously be deformed into a sphere[52].

### Methods
We generate via the molecular-dynamics-based Donev-Stillinger-Torquato (DST) method[53] high-quality strictly jammed packing configurations and then analyze the characteristics of individual packings in order to deduce a novel organizing principle. The packings of frictionless binary superdisks with $x = n_S/(n_S + n_L) = 2/3$ (where $n_L$ and $n_S$ are respectively the number of large and small particles) and $\alpha = R_S/R_L = 5/7$ (where $R_L$ and $R_S$ are the corresponding semi-axis of large and small particles, respectively). Specifically, small non-overlapping particles are initially placed in the periodic simulation box (fundamental cell) with random positions and orientations. The particles are then given random translational and rotational velocities and their motions follow Newtonian dynamics as they collide elastically and also expand uniformly with an expansion rate $\Gamma$, while the fundamental cell deforms to better accommodate the configuration. Large $\Gamma$ values (~0.01) are used in the early stage of the simulation to suppress the formation of order; and very small $\Gamma$ values (~$10^{-6}$) are used when the system is close to jamming in order to establish well-defined contact networks. The final packings are verified to be strictly jammed using the "infinitesimal shrinkage" method with deformable boundary[47].

### Acknowledgements

J.T. gratefully acknowledges the support of Grant NSFC No. 11274200 and NSFSD No. ZR2011AM017. Y.J. thanks Arizona State University for his start-up fund. S.T. was supported in part by the National Science Foundation under Grants No. DMR-0820341 and No. DMS-1211087.

### Author Contributions

J.T., Y.J. and S.T. conceived the research and devised the method, J.T., Y.J., Y.X. and S.T. performed analysis, J.T., Y.J. and S.T. wrote the paper.

### Additional Information

**Supplementary information** accompanies this paper at http://www.nature.com/srep

**Competing financial interests:** The authors declare no competing financial interests.

**How to cite this article**: Tian, J. *et al.* A Geometric-Structure Theory for Maximally Random Jammed Packings. *Sci. Rep.* **5**, 16722; doi: 10.1038/srep16722 (2015).